# Dimension-Dependent Stimulated Radiative Interaction of a Single Electron Quantum Wavepacket

Avraham Gover, Yiming Pan

*Department of Electrical Engineering Physical Electronics,*

*Tel Aviv University, Ramat Aviv 69978, ISRAEL*

**Abstract**

In the foundation of quantum mechanics, the spatial dimensions of electron wavepacket are understood only in terms of an expectation value – the probability distribution of the particle location. One can still inquire how the quantum electron wavepacket size affects a physical process. Here we address the fundamental physics problem of particle-wave duality and the measurability of a free electron quantum wavepacket. Our analysis of stimulated radiative interaction of an electron wavepacket, accompanied by numerical computations, reveals two limits. In the quantum regime of long wavepacket size relative to radiation wavelength, one obtains only quantum-recoil multiphoton sidebands in the electron energy spectrum. In the opposite regime, the wavepacket interaction approaches the limit of classical point-particle acceleration. The wavepacket features can be revealed in experiments carried out in the intermediate regime of wavepacket size commensurate with the radiation wavelength.

**Introduction**

When interacting with a radiation wave under the influence of an external force, free electrons can emit radiation spontaneously, or be stimulated to emit/absorb radiation and get decelerated/accelerated. Such an interaction can also be facilitated without an external force when the electron passes through polarizable medium. Numerous spontaneous radiative emission schemes of both kinds are well known: Synchrotron radiation, Undulator radiation, Compton Scattering, Cerenkov radiation, Smith-Purcell radiation, transition radiation [1-6]. Some of these schemes were demonstrated to operate as coherent stimulated radiative emission sources, such as Free Electron Lasers (FEL) [7-9], as well as accelerating (stimulated absorption) devices, such as Dielectric Laser Accelerator (DLA) and Inverse Smith-Purcell effect [10-12].

All of these spontaneous and stimulated radiation schemes have been analyzed in the classical limit - where they are modeled as point particles, and in the quantum limit - where they are normally modeled as plane waves [13-16]. Semi-classical wavepacket analysis of Kapitza-Dirac scattering was presented in [17]. However, a comprehensive quantum analysis of stimulated radiative interaction of a <u>free electron wavepacket</u> (radiative emission/absorption or equivalently acceleration/deceleration) in a finite interaction length is not available yet. It is required for bridging the classical "point particle" theory of accelerators and free electron radiators (FEL, DLA) with the quantum plane-wave limit theories of such devices, and of related important effects as multiphoton emission/absorption quantum–recoil spectrum in Photon-Induced Near-field Electron Microscopy (PINEM) [18-22].

The interpretation and the essence of the electron quantum wavepacket and its electromagnetic interactions have been a subject of debate since the early conception of quantum mechanics[43]. Modern QED theory and experiments indicate that spontaneous emission by a free electron is independent of its wavepacket dimensions [23-30]. However, in the present paper we focus on the stimulated emission process, and show that in this case the wavepacket dimensions do affect the interaction in a certain range of operation that we define.

In the following we analize the stimulated radiative interaction of a single-electron wavepacket of arbitrary size, establishing first the consistency of our analysis with previous theory and experimental measurements of PINEM, FEL and DLA. We then present the main result: derivation of a new "phase-dependent" stimulated radiative interaction regime of an electron wavepacket in which the physical significance of the wavepacket size and the history of its generation and transport to the interaction region are exhibited. We demonstrate then how the quantum wavepacket theory evolves to the classical point-particle interaction limit in this regime.

**Modeling and Methods**

**First Order Perturbation Analysis** Our one-dimensional interaction model is based on the first order perturbation solution of the relativistic "modified Schrödinger equation", derived from Klein-Gordon equation for the case when the spin effect is negligible [13]( see Supplementary 1):

$$i\hbar \frac{\partial \psi(z,t)}{\partial t} = \left( H_0 + H_I(t) \right) \psi(z,t),$$ (1)

where $H_0 = \varepsilon_0 + v_0 \left( -i\hbar\nabla - \mathbf{p_0} \right) + \frac{1}{2m^*} \left( -i\hbar\nabla - \mathbf{p_0} \right)^2$ is the free space Hamiltonian, $m^* = \gamma_0^3 m$, and the interaction part is:

$$H_I(t) = -\frac{e \left( \mathbf{A} \cdot (-i\hbar\nabla) + (-i\hbar\nabla) \cdot \mathbf{A} \right)}{2\gamma_0 m}$$ (2)

This model is fitting for discription of the variety of interaction schemes mentioned, some of them operating with a relativistic beam. In the present one-dimensional model of electron interaction in a "slow-wave" structure we use a longitudinal vector potential $\mathbf{A} = -\frac{1}{2i\omega} \left( \tilde{\mathbf{E}}(z)e^{-i(\omega t + \phi_0)} - \tilde{\mathbf{E}}^*(z)e^{i(\omega t + \phi_0)} \right)$, where $\tilde{\mathbf{E}}(z) = E_0 e^{iq_z z} \hat{\mathbf{e}}_\mathbf{z}$ represents the dominant slow component of the radiation wave, and transverse field components, as well as transverse variation of the field are neglected. We examplify our modeling here for a case of Smith-Purcell radiation (see Fig. 1), for which the radiation wave is a Floquent mode:

$\tilde{\mathbf{E}}(z) = \sum_m \tilde{\mathbf{E}}_m e^{iq_{zm}z}$ with $q_{zm} = q_{z0} + m2\pi/\lambda_G$ , $\lambda_G$ is the grating period, $q_{z0} = q\cos\Theta$, $q = \omega/c$ and $\Theta$ is the incidence angle of the radiation wave relative to the axial interaction dimension. The radiation wave number $q_z = q_{zm}$ represents one of the space harmonics m that satisfies synchronizm condition with the electron [6]: $v_0 \cong \omega/q_{zm}$. We note that the analysis would be similar for the Cerenkov interaction scheme with $q_z = n(\omega)\cos\Theta$ , and $n(\omega)$ the index of refraction of the medium. Furthermore, the analysis can be extended to the case of FEL and other interaction schemes [13,14].

The solution of Schrodinger equation to zero order (i.e. free-space propagation) is well known. Assuming that the initial wavepacket, which is emitted at some point $z = -L_D$ near the cathode face (or any other electron source) at time $t = -t_D$, is a gaussian at its waist, then:

$$\psi^{(0)}(z,t) = \left(2\pi\sigma_{p_0}^2\right)^{-\frac{1}{4}} \int \frac{dp}{\sqrt{2\pi\hbar}} \exp\left(-\frac{(p-p_0)^2}{4\sigma_{p_0}^2}\right) e^{ip(z+L_D)/\hbar} e^{-iE_p(t+t_D)/\hbar} = \int dp\, c_p^{(0)} e^{-iE_p t/\hbar} |p(z)\rangle, \quad (3)$$

where $|p(z)\rangle \equiv e^{ipz/\hbar}/\sqrt{2\pi\hbar} |p\rangle$, $L_D$, $t_D = L_D/v_0$ are the "effective" drift length and drift time of the wavepacket center from a "virtual cathode". Expanding the energy dispersion relation to second order $E_p = c\sqrt{m^2c^2 + p^2} \approx \varepsilon_0 + v_0(p - p_0) + \frac{(p-p_0)^2}{2m^*}$, the wavepacket development in momentum space is then given by (see supplementary 2):

$$c_p^{(0)} = \left(2\pi\sigma_p^2\right)^{-\frac{1}{4}} \exp\left(-\frac{(p-p_0)^2}{4\tilde{\sigma}_p^2(t_D)}\right) e^{i(p_0 L_D - E_0 t_D)/\hbar}, \qquad (4)$$

with $\tilde{\sigma}_p^2(t_D) = \sigma_{p_0}^2\left(1 + it_D/t_{R_\parallel}\right)^{-1}$, $\sigma_{p_0} = \hbar/2\sigma_{z_0}$, $t_{R_\parallel} = \frac{m^*\hbar}{2\sigma_{p_0}^2} = 4\pi\frac{\sigma_{z_0}^2}{\lambda_c^* c}$, and we defined $\lambda_c^* = \lambda_c/\gamma^3$ with $\lambda_c = h/mc$ – the Compton wavelength.

Note that the "virtual cathode" is not necessarily the physical face of the electron beam source. We define it as the point where the electron wavepacket is at its minimal length (at its waist). Recent work showed that this position can be optically controlled by streaking techniques [37].

We now solve Eq.1 in the interaction region $0<z<L_I$ using the first order perturbation theory in momentum space (see Supplementary 2)

$$\psi(z,t) = \psi^{(0)}(z,t) + \psi^{(1)}(z,t) = \int dp \left( c_p^{(0)} + c_p^{(1)} \right) e^{-iE_p t/\hbar} \left| p(z) \right\rangle. \tag{5}$$

and then calculate the electron momentum density distribution after interaction:

$$\rho(p') = \rho^{(0)}(p') + \rho^{(1)}(p') + \rho^{(2)}(p')$$
$$= \frac{\left| c^{(0)}(p') \right|^2 + 2\,\mathrm{Re}\left\{ c^{(1)*}(p') c^{(0)}(p') \right\} + \left| c^{(1)}(p') \right|^2}{\int dp' \left( \left| c^{(0)}(p') \right|^2 + \left| c^{(1)}(p') \right|^2 \right)}, \tag{6}$$

where

$$\rho^{(0)}(p') = \left| c^{(0)}(p') \right|^2 = \left( 2\pi\sigma_{p_0}^2 \right)^{-\frac{1}{2}} \exp\left( -\frac{(p'-p_0)^2}{2\sigma_{p_0}^2} \right) \tag{7}$$

is the initial Gaussian momentum density distribution.

## Results

**Phase-Independent Momentum Distribution – Fel Gain.** First we draw attention to the second order density distribution (third term in Eq.6), as derived in Supplementary 2:

$$\rho^{(2)}(p') = \Upsilon^2 \left[ \left( \frac{p' + p_{rec}^e/2}{p_0} \right)^2 \rho^{(0)}(p' + p_{rec}^e) - \left( \frac{p_0 - p_{rec}^{(e)}/2}{p_0} \right)^2 \rho^{(0)}(p') \right] sinc^2\left( \frac{\bar{\theta}_e}{2} \right)$$
$$+ \Upsilon^2 \left[ \left( \frac{p' - p_{rec}^a/2}{p_0} \right)^2 \rho^{(0)}(p' - p_{rec}^a) - \left( \frac{p_0 + p_{rec}^{(a)}/2}{p_0} \right)^2 \rho^{(0)}(p') \right] sinc^2\left( \frac{\bar{\theta}_a}{2} \right). \tag{8}$$

where the two terms in $\rho^{(2)}$ display two spectral sidebands proportional to the initial density distribution $\rho^{(0)}$ shifted centrally to $p_0 \mp p_{rec}^{(e,a)}$ due to photon emission and absorption recoils (see Fig.2). Its first order moment results in the momentum acceleration/deceleration associated with stimulated radiative interaction. The detailed intergration was carried out in Sup. 3 with a Gaussian momentum distribution (7), and in the limit $p_{rec}^{e,a}, \sigma_{p_0} \ll p_0$ it results in:

$$\Delta p^{(2)} = \int \rho^{(2)}\left(p'\right) p' dp' = \Upsilon^2 \left(\frac{\hbar\omega}{v_0}\right)\left[sinc^2\left(\frac{\bar{\theta}_a}{2}\right) - sinc^2\left(\frac{\bar{\theta}_e}{2}\right)\right]. \qquad (9)$$

where $\Upsilon = \dfrac{eE_0 L_I}{2\hbar\omega}, \bar{\theta}_{e,a} = \bar{\theta} \pm \dfrac{\varepsilon}{2}$, and $\bar{\theta} = \left(\dfrac{\omega}{v_0} - q_z\right)L_I$ is the classical "interaction detuning parameter"[9], and $\varepsilon = \delta\left(\dfrac{\omega}{v_0}\right)L_I = 2\pi\delta L_I/\beta_0\lambda$ is the interaction–length quantum recoil parameter[13] and $\delta = \hbar\omega/2m^*v_0^2 \ll 1$ . Remarkably Eq.9 is independent of the wavepacket distribution $\rho^{(0)}$ and the wavepacket size, and is satisfactorily consistent with the stimulated emission/absorption terms in the quantum-electrodynamic photon emission rate expression $dv_q/dt$, that was derived in [13] for a single plane-wave electron wavefunction in the limit $\rho^{(0)}\left(p'\right) = \delta\left(p - p'\right)$:

$$\frac{dv_q}{dt} = \Gamma_{sp}\gamma_0 v_0\left[\left(v_q + 1\right)\frac{1}{\gamma_e v_e}sinc^2\left(\frac{\bar{\theta}_e}{2}\right) - v_q \frac{1}{\gamma_e v_e}sinc^2\left(\frac{\bar{\theta}_a}{2}\right)\right]. \qquad (10)$$

Using Eq. 9 and the stimulated emission part of Eq. 10 in a conservation of energy and momentum relation,

$$\Delta p^{(2)} = -\frac{L_I}{v_0}\left(\frac{\hbar\omega}{v_0}\right)\left(\frac{dv_q}{dt}\right)_{st}, \qquad (11)$$

we get an interesting relation between the interaction parameter $\Upsilon$ and the spontaneous radiation emission coefficient $\Gamma_{sp}$

$$\Upsilon^2 = \frac{L_I}{v_0}\Gamma_{SP}v_q, \qquad (12)$$

This explicitly relates the QED terms of spontaneous photon emission rate per mode ($\Gamma_{SP}$) and the incident photon number flow into the mode ($v_q$) in stimulated emission, to the semi-classical parameter of the interacting field component $|E_0|$. It is also interesting to point out that in the limit of negligible recoil relative to the finite-length "homogeneous broadening" emission/absorption lines, Eq. 9 reduces to:

$$\varepsilon = \pi \frac{\hbar\omega}{m^*v_0^2} \frac{L_1}{\beta_0\lambda} \ll 1, \tag{13}$$

$$\Delta p^{(2)} = \frac{\hbar\omega}{v_0} \Upsilon^2 \varepsilon \frac{d}{d\theta} \operatorname{sinc}^2(\bar{\theta}/2), \tag{14}$$

consistently with the conventional classical gain expression of Smith-Purcell and Cerenkov FELs, as well as other FELs in the low-gain regime[9,13].

It should be noted that second order (in the field) acceleration/deceleration (Eq. 9) is only possible out of synchronism ($\bar{\theta} \neq 0$) and is due to the asymmetry of emission and absorption recoils in the interaction length. Net acceleration/deceleration (gain/loss) is possible in the FEL quantum limit $\varepsilon \gg 1$ (opposite of Eq.13) in which the homogeneously broadened emission/absorption lines do not overlap (Eq.9-10). In the more common case of classical FEL gain (Eq.14): $\bar{\theta}_e \simeq \bar{\theta}_a \simeq \bar{\theta}$, gain/attenuation are possible only if the degeneracy of emission/absorption is lifted by operating out of synchronism ($\bar{\theta} \neq 0, \sim \pi$).

In PINEM, the near field interaction takes place along a very short interaction length relative to the wavelength. Our analysis reduces to this case when $\varepsilon = 2\pi L_1 / \beta_0\lambda \ll \pi$ and $p_{rec}^{(e)} \simeq p_{rec}^{(a)} \simeq p_{rec}^{(0)}$. In this case, the emission and absorption lines in the spectrum $\rho^{(2)}$ (Eq. 8) are degenerate and symmetric around $p_0$, and there is no net-gain/acceleration. However, the emission and absorption lines in the momentum distribution function are still separable then, if the quantum recoil momentum $p_{rec}^{(0)}$ is significant relative to the wavepacket momentum spread

$$p_{rec}^{(0)} = \frac{\hbar\omega}{v_0} \gg \sigma_{p0},$$

(15)

In this limit (8) results in:

$$\rho = \rho^{(0)} + \rho^{(2)} \simeq \left(1 - 2\Upsilon^2\right)\rho^{(0)}\left(p'\right) + \Upsilon^2\left[\rho^{(0)}\left(p' + p_{rec}^{(0)}\right) + \rho^{(0)}\left(p' - p_{rec}^{(0)}\right)\right]$$

(16)

representing symmetric sidebands, spaced by $p_{rec}^{(0)}$ on both sides of the center initial momentum $p_0$ as displayed in Fig. 2. Similar sideband development is shown in supplementary video 1 based on numerical solution of Eq. 1 for weak interaction $\Upsilon < 1$. This result is similar to the measured spectrum in PINEM experiments [18-22], where multiple ($\gg 3$) sidebands were observed due to multiple photon emission/absorption. In our example of weak interaction, only two emission/absorption sidebands are observed. Note that also the fundamental sideband has a second order reduction factor originating from the important renormalization denominator of Eq. 6 (see the reduction effect of $\rho^{(0)}$ in Fig. 2). Note that contrary to the symmetric momentum distribution spectrum (Eq.16), characteristic to PINEM, in the limit of long interaction length $\varepsilon = 2\pi\delta L_q/\beta_0\lambda \gg 1$, which is characteristic to the quantum FEL[13,14,39], the energy spectrum (8) depends on the synchronizm detuning, exhibiting net deceleration or acceleration corresponding to detuning to $\bar{\theta}_e \simeq 0$ and $\bar{\theta}_a \simeq 0$ respectively (Eq. 9).

**Phase-Dependent Momentum Distribution.** The second order perturbation term of the momentum distribution lost the dependence on the phase $\phi_0$ of the wavepacket center relative to the laser field, and <u>does not reveal any specific features of the single electron wavepacket</u>. We now draw attention to the phase-dependent density distribution (Supplementary Eq.32). This term has not been considered previously in the literature, but in the present work it is of prime interest because it retains the dependence on the phase $\phi_0$. In the limit of negligible recoil parameter $\varepsilon = 0, \delta = 0$ and to first order in $p_{rec}^{(0)}$, one obtains

$$\rho^{(1)}\left(p'\right) = 2\left(\frac{\hbar\omega/v_0}{p_0}\right)\Upsilon\rho^{(0)}\left(p'\right)sinc\left(\frac{\bar{\theta}}{2}\right)e^{-\Gamma^2/2}cos\left(\frac{\bar{\theta}}{2} + \phi_0 + \varphi\right)\left[1 - \frac{\left(p' - p_0\right)p'}{\sigma_{p_0}^2}\right].$$

(17)

where

$$\Gamma = \frac{\omega}{v_0} \sigma_z (t_D) = \frac{2\pi\sigma_z (t_D)}{\beta\lambda},$$

$$\varphi = \frac{2\omega t_D}{mv_0} (p' - p_0).$$

(18)

The coefficient $e^{-\Gamma^2/2}$ assures that the first order term of the momentum distribution $\rho^{(1)}$ is diminished when $\Gamma \gg 1$, namely when the wavepacket expands on its way from the source beyond the size of the interaction wavelength. Neglecting $\varphi$, the momentum transfer is:

$$\Delta p^{(1)} = \int \rho^{(1)} (p') p' dp' = -\left( \frac{eE_0 L_t}{v_0} \right) e^{-\Gamma^2/2} sinc\left( \frac{\bar{\theta}}{2} \right) cos\left( \phi_0 + \frac{\bar{\theta}}{2} \right),$$

(19)

for $\Upsilon \ll 1$. This expression can be contrusted with the classical "point-particle" momentum transfer equation (see Eq. 11 in the Supplementary)

$$\Delta p_{point} = -\frac{eE_0 L_t}{v_0} sinc\left( \frac{\bar{\theta}}{2} \right) cos\left( \phi_0 + \frac{\bar{\theta}}{2} \right).$$

and the corresponding expression for stimulated-superradiant emission energy $W_q = v_0 \Delta p_{point}$, that would be expected from conservation of energy[31]. Except for the reduction factor:

$$\Delta p^{(1)} / \Delta p_{point} = e^{-\Gamma^2/2},$$

(20)

the acceleration/deceleration of the quantum wavepacket (Eq.19) scales with $\bar{\theta}$ and $\phi_0$ similarly to a point particle.

Figures 3 and 4 display the momentum density distribution $\rho^{(0)} + \rho^{(1)}$ for the case $\bar{\theta} = 0$ and acceleration phase: $\phi_0 = 0$. In this explicitly quantum limit of wavepacket acceleration the resultant post-interaction distribution $\rho(p') = \rho^{(0)} + \rho^{(1)} + \rho^{(2)}$ does not change much in width, but it gets lopsided towards positive momentum. The momentum distribution is shifted about $\Delta p^{(1)} = \dfrac{2\Upsilon}{e}\dfrac{\hbar\omega}{v_0}$ for $\Gamma = \sqrt{2}$. A left shifted momentum distribution would appear in the case of deceleration for phase $\phi_0 = \pi$. These low-gain phase dependent acceleration/deceleration features are comfirmed by the numerical simulations (Supplementary video-2(a,b)).

**Discussions**

**The Wave-Particle Transition.** The scaling of the first order perturbation expression (19) with wavepacket-size $\sigma_z(L_D)$ (Eq. 18) that was derived in our comprehensive semiclassical analysis, demonstrates successfuly the transition of the quantum wavepacket regime to the point-particle regime. In the long quantum wavepacket limit $\Gamma(L_D) = 2\pi\sigma_z(L_D)/\beta\lambda \gg 1$ the wavepacket behaves as a plane wave and no phase-dependent synchronous wave acceleration is possible, and therefore $\Delta p^{(1)}/\Delta p_{point}$ diminishes exponentialy. In the oposite limit the wavepacket acceleration reduces to the phase-dependent point-particle acceleration characteristics.

To support our first order perturbation analytical study of the wave/particle transition, we employed numerical solution of Eq. 1 for different values of $\Upsilon$, starting with initial conditions of a wavepacket, entering the interaction region with different values of its size $\sigma_z(L_D)$. Videos 2, 3 display phase-dependent linear acceleration/deceleration in the intermediate wavepacket parameters regime of wave/particle transition: $\Gamma = 0.6$. The parameters of the Video2 example are in the small field case $\Upsilon = 0.2$, where the small momentum distribution shift resembles the analytical result of Figs. 3,4, and video3 coresponds to strong field $\Upsilon = 1.5$, displaying a wavepacket, accelerating almost as a point particle. Comparison of Videos 2 and 3 indicates near proportionality relation of the net acceleration with $\Upsilon$(or $E_0$). Note that the measurable resolution of momentum transfer relative to the wavepacket momentum spread is limited by the inequality

$$\frac{\Delta p}{\sigma_{po}} < \frac{\Delta p_{point}}{\sigma_{p0}} = 8\pi\Upsilon\frac{\sigma_{z0}}{\lambda\beta} < 8\pi\Upsilon\Gamma, \tag{21}$$

and therefore, resolving the momentum spectrum shift requires $\Upsilon$ large enough, not to be limited by this inequality. The numerical simulation example (supplementary video 3) displays such resolved incremental momentum shift of the accelerated wavepacket spectrum. Analytical description of such parameters example would have required higher order perturbation (multiphoton exchange) analysis [14, 38] beyond the present first order perturbation analysis.

In Fig. 5 we show the computed dependence of $\Delta p / \Delta p_{point}$ on $\Gamma(L_D)$ in the transition range $0.2 < \Gamma < 3$. The two computed examples of $\Upsilon = 0.4, 1.6$ follow quite accurately a Gaussian curve $\exp(-\Gamma^2/2)$, that decays for $\Gamma \gg 1$, and approaches the limit $\Delta p / \Delta p_{point} \to 1$ for $\Gamma \ll 1$, as in Eq.20. The numerical solution is valid beyond the first order perturbation theory and shows good match with the analytical solution in the parameter range $\Upsilon < \sim 1$. Remarkably, the theory and all examples of numerical computations shown in Fig. 5 fit well the Gaussian dependence on the parameter $\Gamma(L_D) = 2\pi\sigma_z(t_D)/\beta\lambda$ -confirming that the relevant wavepacket parameter $\sigma_z(t_D)$ is the <u>history-dependent</u> spatial size of the wavepacket after drift time $t_D$ , independently of the initial wavepacket size $\sigma_{z0}$ .

An important observation can be made now based on this conclusion. From inspection of the expression for the wavepacket expansion in free drift:

$$\sigma_z\left(t_D\right) = \sqrt{\sigma_{z0}^2 + \left(\frac{\lambda_c^* \, ct_D}{4\pi \, \sigma_{z_0}}\right)^2} \, , \tag{22}$$

it is obvious that $\sigma_{z_0} \to 0$ is not the point particle limit, since then the wavepacket size exploads, and the absolute minimum of $\sigma_z\left(t_D\right)$ <u>for any given</u> $t_D$ , $(or \; L_D = v_0 t_D)$ is:

$$\sigma_z\left(t_D\right)\big|_{\min} = \sqrt{\frac{\lambda_c^*}{2\pi} ct_D} \, , \tag{23}$$

This corresponds to a minimum value of $\Gamma$ for fixed freqency $\omega$ , which means that for fixed drift length $L_D$ the curve in Fig. 5 has physical meaning only for $\Gamma > \Gamma_{min} = \frac{\omega}{v_0} \sigma_z (t_D) \Big|_{min}$ . We define a critical drift length $z_G = v_0 t_D$ as the distance for which $\Gamma_{min} = \sqrt{2}$, that corresponds to the point where $\Delta p / \Delta p_{point} = 1/e$ :

$$z_G = \frac{\beta_0^3 \gamma_0^{\ 3}}{\pi} \frac{\lambda^2}{\lambda_c},$$

(24)

We come to the significant observation that for drift distances away from the source $L_D \gg z_G$, wavepacket-dependent linear (in the field) accleration/deceleration of a single electron is always diminished. This observation is also consistent with an earlier suggestion that the quantum wavepacket spread poses a fundamental physical high frequency (or short wavelength limit - $\lambda_{cutoff} = (\pi \lambda_c / \beta^3 \gamma^3)^{1/2}$) on measurement of particle beam shot-noise, challeging the conventional mathematical "point-particle" model presentation of shot-noise as an unbound "white noise" [32].

**Wavepacket-Dependent Stimulated Interaction Measurement** The phase-dependent linear field acceleration regime of a wavepacket $\left( \rho^{(1)}, \Delta p^{(1)} \right)$ is of fundamental interest, because contrary to the phase-independent acceleration $\left( \rho^{(2)}, \Delta p^{(2)} \right)$, its characteristics depend on the quantum wavepacket dimensions, and at the same time respond to the classical phase of the wave. So far, previous finite interaction length laser-acceleration experiments of single electrons were carried out in the classical regime [10-12],

We point out that the effect of the wavepacket size on the acceleration of a single electron wavepacket can be measured in the phase-dependent interaction regime by varying the the parameter $\Gamma = 2\pi \sigma_z (t_D) / \beta_0 \lambda$ over a range around $\Gamma < \approx \sqrt{2}$ , and measuring a wavepacket acceleration dependence as in Fig. 5 and laser-phase ($\phi_0$) dependence as in Eq.18. Experimentally, for a fixed electron source, one can scan over $\Gamma(t_D)$ by varying the interaction wavelength or beam velocity - $\lambda, \beta_0$ , or performing the radiative interaction at different drift distances $L_D = v_0 t_D$ within the range $L_D < \approx z_G (\lambda, \beta_0)$ Eq. 24).

Alternatively, as recently demonstrated [40], it is possible to control the waepacket parameters and size $\sigma_z(t_D)$, by use of optical streaking techniques.

A measurement of acceleration at different distances and wavelengths is analogous to measurement of the transverse diffraction characteristics of a laser beam of wavelength λ by measuring its transmission through apertures of radius R at different distances L from the beam waist. Such measurements reveal information about the beam, if made at distances shorter than or commensurate with the aperture Rayleigh length $z_R = \pi R^2 / \lambda$. At distances $z > z_R$ the transmitted beam would always be attenuated, independently of the initial waist size, while for $z < z_R$ the transmission through the aperture depends on the spot size at the aperture. Analogously, in the stimulated emission experiments the radiation interaction wavelength $\beta_0 \lambda$ acts as a moving longitudinal aperture (analogous to the aperture R) with the longitudinal Compton wavelength $\lambda_c^* / \beta$ playing the same role as the optical wavelength λ in the diffraction analogue.

**Measurement Limits** One may want to use momentum (energy) distribution measurement after the electron radiative interaction in the phase-dependent wavepacket transition regime, in order to determine the quantum-wavepacket dimensions of electrons emitted from natural electron emission sources. However, measurement of the electron momentum distribution requires normally accumulation of data from an ensemble of particles emitted from the electron source. Measurements of the wavepacket characteristics of electrons, photo-emitted from single-electron emission sources like a tip [34] or an ion cold-trap [35], would be usually masked by random (thermal) spread of the particles in the ensemble (see discussion in supplementary). This leads to "inhomogeneous broadening" of the e-beam energy spectrum due to the ensemble momentum spread (see [14, 39]). After a short drift length, the classical statistical spatial spread of the ensemble exceeds the size of the individual wavepackets [36]. This limits the practical range within which the stimulated interaction experiment would reflect the intrinsic spatial dimension of the single electrons. In order to measure the single electron wavepacket dimension, the electron beam must be energy-filtered and preselected in time, and the measurement would be essentially a protected weak measurement of the Aharonov-Vaidman's kind [40].

**Conclusions**

Here we related to the fundamental questions of the physical significance of a single particle quantum wavepacket and the wave-to-particle transition to the classical electrodynamics limit. The presented semiclassical first order perturbation analysis and the numerical computations show that stimulated finite-length interaction of a single electron wavepacket with radiation can be dependent on the features of the wavepacket, its history and its spatial position relative to the phase of the accelerating radiation wave. This can only happen within a certain range, close enough to the electron emission source $L_D <\approx z_G$ (24). It was shown, that contrary to matter-wave interference approach[41,42], the stimulated radiative interaction measures the evolving electron wavepacket size $\sigma_z(t_D)$, and not the intrinsic "coherence length" $\sigma_{z0}$.

On passing, we also showed that the phase-independent, second order (in the field) acceleration solution of stimulated radiative interaction of an electron wavepacket is consistent with the quantum limit of PINEM and the quantum and classical multi-particle theory limits of DLA and FEL.

**Acknowlegements**

We acknowledge Yakir Aharonov, Aharon Friedman, Shlomo Rushin, Amnon Yariv and Wolfgang Schleich for useful discussions and comments. Especially, we thank Peter Kling for pointing out an earlier mistake in derivation. The work was supported in parts by DIP (German-Israeli Project Cooperation) and US-Israel Binational Science Foundation, and by the PBC program of the Israel council of higher education. Correspondance and requests for materials should be addressed to Y.P. (yimingpan@mail.tau.ac.il) and A. G.(gover@eng.tau.ac.il).

**Figures**

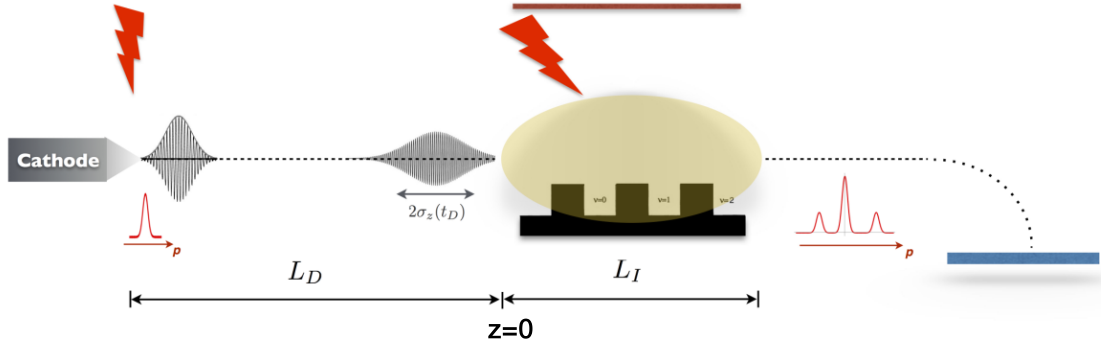

Fig. 1: The experiment setup. Single electron wavepackets are photo-emitted from a cathode driven by a ftSec laser. After a free propagation length $L_D$, the expanded wavepacket passes next to the surface of a grating , and interacts with the near-field radiation, that is excited by an IR wavelength laser, phase locked to the photo-emitting laser. The momentum distribution of the modulated wavepacket is measured with an electron energy spectrometer.

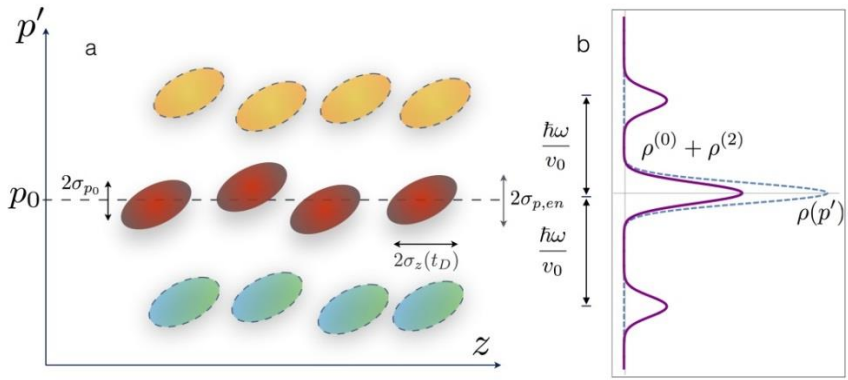

Fig. 2: The quantum recoil limit of electron-laser interaction. The phase-space distribution of an ensemble of quantum wavepackets after interaction with the near-field is shown in (a), and its final momentum distribution is shown in (b). In this limit the condition $p_{rec} = \dfrac{\hbar\omega}{v_0} \gg \sigma_{en} > \sigma_{p0}$ is satisfied, where $\sigma_{en}$ is the ensemble momentum spread.

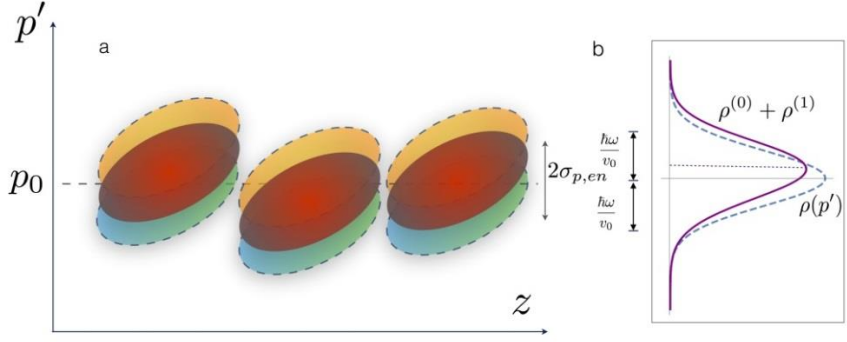

Fig. 3: Linear field acceleration of a phase-defined wavepacket. The phase space distribution of electron quantum wavepackets interacting weakly with the near-field wave is shown in (a), and its final momentum distribution in (b). In this limit $\Gamma = \dfrac{2\pi\sigma_z\left(t_D\right)}{\beta\lambda} \simeq 1$ and $p_{rec} = \dfrac{\hbar\omega}{v_0} << \sigma_{p0}$.

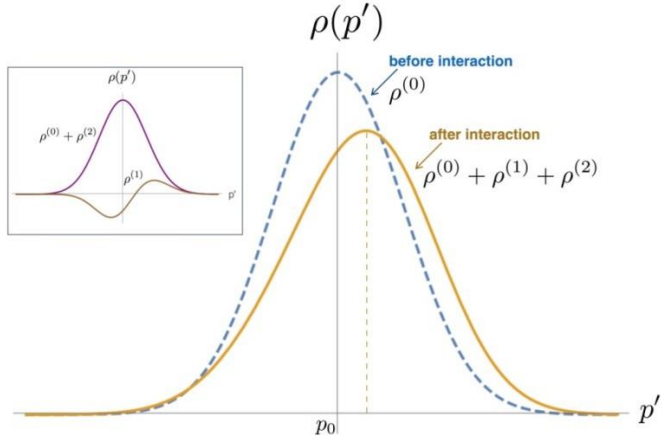

Fig. 4: Total final momentum distribution of a Linear field accelerated phase-defined wavepacket. The parameters are in the quantum-classical transition range: $\Gamma = \sqrt{2}$ for maximal momentum gain: $\bar{\theta} = 0$ (velocity synchronizm), $\phi_0 = 0$ (accelerating phase). The inset shows the incremental distributions $\rho^{(1)}$ and $\rho^{(0)} + \rho^{(2)}$ seperately.

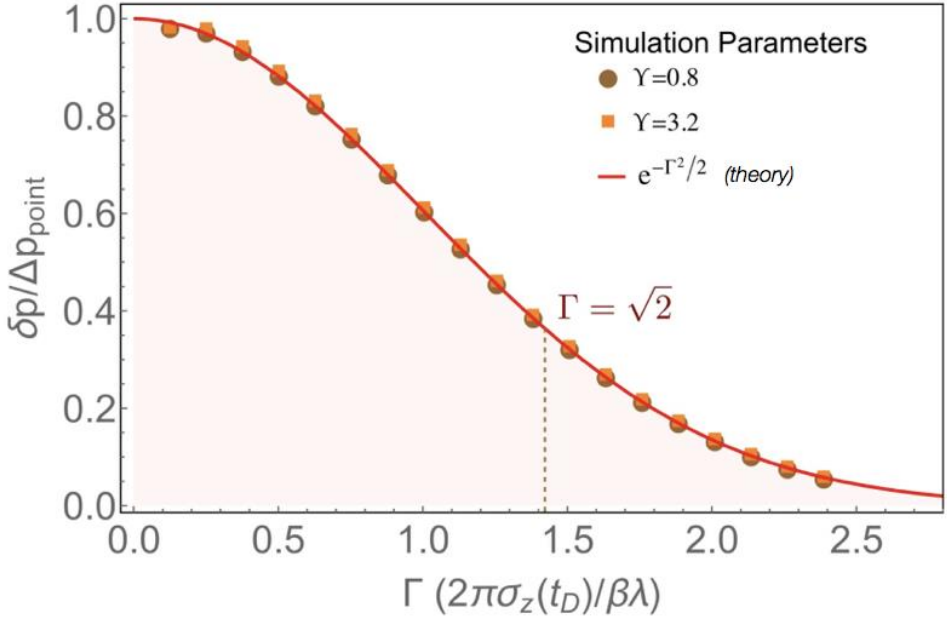

Fig. 5: The reduction factor of wavepacket acceleration relative to the "point-particle" classical case, as calculated from numerical solution of Schodinger equation (points), and compared to the first-order perturbation expression (eq.19) (red curve). The numerical computation with parameters $\beta_0 = 0.7, \gamma_0 = 1.4, \beta_0\lambda = \Lambda_G = 2\mu m, L_I = 8\mu m$ reproduces the classical (point-particle) limit $\delta p / \Delta p_{point} \rightarrow 1$ for two different interaction-strength simulation parameters: $\Upsilon = 0.8$ and 3.2.

# Supplementary Materials:

## 1. The modified 'relativistic' Schrodinger equation from the Klein-Gordon equation

Some of the electron-radiation interaction schemes referred to in the paper (Smith-Purcell radiation, PINEM, FEL etc.) operate with a relativistic beam, therefore the use of Schrodinger equation would not be satisfactory for all cases of interest. Since spin effects are not relevant for the present problem, we do not need to use Dirac equation, but rather base our analysis on the Klein-Gordon (KG) equation, that is directly derived from the Dirac equation. Furthermore, following Ref. 13, we reiterate the derivation of a Schrödinger-like equation out of the KG equation, using second order iterative expansion of the free electron energy around its center energy $\varepsilon_0 = \sqrt{p_0^2 c^2 + m^2 c^4}$. This expansion reduces the quadratic KG equation into the parabolic Schrodinger equation under the well-satisfied approximation that the initial momentum spread and the momentum change due to the interaction are within a range $\Delta p \ll p_0$.

The Klein-Gordon equation originates from the relativistic energy-momentum dispersion:

$$E_p^2 = p^2 c^2 + m^2 c^4, \tag{1}$$

where m is the electron rest mass and c the speed of light. To obtain the KG equation, we replace $E \rightarrow i\hbar \frac{\partial}{\partial t}, \mathbf{p} \rightarrow -i\hbar\nabla - e\mathbf{A}$ （minimal coupling with electromagnetic radiation） and apply the differential operator on a wavefunction:

$$\left(i\hbar\frac{\partial}{\partial t}\right)^2 \psi\left(r,t\right) = c^2 \left(-i\hbar\nabla - eA\right)^2 \psi\left(r,t\right) + m^2 c^4 \psi\left(r,t\right) \tag{2}$$

where e is an electron charge. The KG equation can describe the relativistic electrons with most radiation schemes considered, if spin effect is negligible. If the radiation field is weak, $e\mathbf{A}/mc \ll 1$, then excitation of the negative (positron) energy brunch of the dispersion equation is negligible and one can approximate the wavefunction $\psi\left(r,t\right)$ with a single quasi-harmonic positive energy wave [Ref.13 in the main text]

$$\psi\left(r,t\right)=u\left(r,t\right)e^{-i\varepsilon_{p_0}t/\hbar} \tag{3}$$

where $\varepsilon_0=\sqrt{p_0^2c^2+m^2c^4}=\gamma_0mc^2$, $p_0$ the center momentum and $u\left(r,t\right)$ is a slowly varying function of time. Then substitution of Eq.(3) in (2) and cancelling the fast-varying coefficient $e^{-i\varepsilon_0t/\hbar}$, results in

$$i\hbar\frac{\partial u\left(r,t\right)}{\partial t}=\left(\frac{c^2\left(-i\hbar\nabla-e\mathbf{A}\right)^2+\left(m^2c^4-\varepsilon_0^2\right)}{2\varepsilon_0}\right)u\left(r,t\right)+\frac{\hbar^2}{2\varepsilon_0}\frac{\partial^2u\left(r,t\right)}{\partial t^2} \tag{4}$$

This is an exact expression for the slow part function $u\left(r,t\right)$. Its first order approximation in the time derivatives is

$$i\hbar\frac{\partial u\left(r,t\right)}{\partial t}=\left(\frac{c^2\left(-i\hbar\nabla-e\mathbf{A}\right)^2+\left(m^2c^4-\varepsilon_0^2\right)}{2\varepsilon_0}\right)u\left(r,t\right)$$

Successive iterative substitution of this equation into the exact formula (4) results in now

$$i\hbar\frac{\partial u\left(r,t\right)}{\partial t}=\left(\frac{c^2\left(-i\hbar\nabla-e\mathbf{A}\right)^2+\left(m^2c^4-\varepsilon_0^2\right)}{2\varepsilon_0}\right)u\left(r,t\right)-\frac{1}{2\varepsilon_0}\left(\frac{c^2\left(-i\hbar\nabla-e\mathbf{A}\right)^2+\left(m^2c^4-\varepsilon_0^2\right)}{2\varepsilon_0}\right)^2u\left(r,t\right)$$

$$\tag{5}$$

Now the Klein-Gordon equation can be re-expressed in the form of a modified Schrodinger equation,

$$i\hbar\frac{\partial\psi\left(r,t\right)}{\partial t}=i\hbar\frac{\partial u\left(r,t\right)}{\partial t}e^{-i\varepsilon_0t/\hbar}+\varepsilon_0u\left(r,t\right)e^{-i\varepsilon_0t/\hbar}=H\psi\left(r,t\right) \tag{6}$$

where the effective Hamiltonian is

$$H=\varepsilon_0+\left(\frac{c^2\left(-i\hbar\nabla-e\mathbf{A}\right)^2+\left(m^2c^4-\varepsilon_0^2\right)}{2\varepsilon_0}\right)-\frac{1}{2\varepsilon_0}\left(\frac{c^2\left(-i\hbar\nabla-e\mathbf{A}\right)^2+\left(m^2c^4-\varepsilon_0^2\right)}{2\varepsilon_0}\right)^2 \tag{7}$$

The Hamiltonian can be split into an electronic unperturbed part and a radiative perturbation part $H = H_0 + H_I(t)$ , where

$$H_0 \simeq \varepsilon_0 + v_0 \left( -i\hbar\nabla - \mathbf{p_0} \right) + \frac{1}{2\gamma_0^3 m} \left( -i\hbar\nabla - \mathbf{p_0} \right)^2 \qquad (8)$$

and to the first order in $\mathbf{A}$ ,

$$H_I(t) = -\frac{e \left( \mathbf{A} \cdot \left( -i\hbar\nabla \right) + \left( -i\hbar\nabla \right) \cdot \mathbf{A} \right)}{2\gamma_0 m} = -\frac{e \left[ \mathbf{A} \cdot \left( -i\hbar\nabla \right) - i\hbar \left( \nabla \cdot \mathbf{A} \right) / 2 \right]}{\gamma_0 m} \qquad (9)$$

where $v_0 = p_0 / \gamma_0 m = \beta_0 c$ . For the case of our concern,

$$\mathbf{A} = -\frac{1}{2i\omega} \left( \tilde{\mathbf{E}}(z) e^{-i(\omega t + \phi_0)} - \tilde{\mathbf{E}}^*(z) e^{i(\omega t + \phi_0)} \right)$$

$$\mathbf{E} = -\frac{\partial \mathbf{A}}{\partial t} = \mathrm{Re}[\tilde{\mathbf{E}}(z) e^{-i(\omega t + \phi_0)}] = E_0 \cos \left( q_z z - \omega t - \phi_0 \right) \qquad (10)$$

In our present one-dimensional analysis, we assume a longitudinal field component of a slow wave structure (e.g. a grating) $\tilde{\mathbf{E}}(z) = E_0 e^{iq_z z} \hat{\mathbf{e}}_z$ , neglecting transverse field components and transverse variation of the field. This derived modified relativistic Scrodinger equation with the effective Hamiltonians (8), (9) are used in the main manuscript for the perturbative solution.

For reference, we briefly derive here the "point-particle" momentum transfer under such a field:

$$\begin{aligned}
\Delta p_{point} &= -e \int_0^{t_f} \mathrm{Re}[\tilde{\mathbf{E}}(z(t)) e^{-i(\omega t + \phi_0)}] dt \\
&= -e \int_0^{t_f} E_0 \cos \left( q_z z(t) - \omega t - \phi_0 \right) dt \\
&\simeq -e E_0 \int_0^{t_f} \cos \left( q_z v_0 t - \omega t - \phi_0 \right) dt \\
&= -\frac{e E_0 L_I}{v_0} sinc \left( \frac{\bar{\theta}}{2} \right) cos \left( \phi_0 + \frac{\bar{\theta}}{2} \right).
\end{aligned} \qquad (11)$$

where classically we assume that $\dot{z} \simeq v_0, \bar{\theta} = \left( \dfrac{\omega}{v_0} - q_z \right) L_I$ and the interaction length is $L_I = v_0 t_I$.

## 2. First order perturbation analysis

The equation to be solved is the 'relativistic' Schrodinger equation

$$i\hbar \frac{\partial \psi(z,t)}{\partial t} = \left( H_0 + H_I(t) \right) \psi(z,t),$$  (12)

with $H_0$ given by Eq.8 and $H_1$ given by Eq.10. We solve this equation by perturbation theory. The solution of Schrodinger equation to zero order (i.e. free space propagation) is well known. Assuming that the initial wavepacket, which is emitted at some point $z = -L_D$ near the cathode face at time $t = -t_D$, is a gaussian at its waist, then:

$$\psi^{(0)}(z,t) = \left(2\pi\sigma_{p_0}^2\right)^{-\frac{1}{4}} \int \frac{dp}{\sqrt{2\pi\hbar}} \exp\left(-\frac{(p-p_0)^2}{4\sigma_{p_0}^2}\right) e^{ip(z+L_D)/\hbar} e^{-iE_p(t+t_D)/\hbar} = \int dp c_p^{(0)} e^{-iE_p t/\hbar} \left| p(z) \right\rangle, \quad (13)$$

where $\left| p(z) \right\rangle = e^{ipz/\hbar} / \sqrt{2\pi\hbar} \left| p \right\rangle$, and at fixed drift time $t_D$ when t=0:

$$c_p^{(0)} = \left(2\pi\sigma_{p_0}^2\right)^{-\frac{1}{4}} \exp\left(-\frac{(p-p_0)^2}{4\sigma_{p0}^2}\right) e^{i(p_0 L_D - E_0 t_D)/\hbar}.$$

Note that $L_D$, $t_D = L_D/v_0$ are the "effective" drift length and drift time of the wavepacket center, that are somewhat different from the geometric distance and drift time from the cathode face. This is because of the initial section of electron acceleration from the cathode and because the wavepacket longitudinal waist may be somewhere within the cathode.

Expanding the energy dispersion relation to second order

$$E_p = c\sqrt{m^2 c^2 + p^2} \approx \varepsilon_0 + v_0 \left(p - p_0\right) + \frac{\left(p - p_0\right)^2}{2m^*}, \quad (14)$$

And substituting in $c_p^{(0)}$, the wavepacket development in momentum space at time $t_D$ can be written as:

$$c_p^{(0)} = \left(2\pi\sigma_{p_0}^2\right)^{-\frac{1}{4}} \exp\left(-\frac{\left(p-p_0\right)^2}{4\tilde{\sigma}_p^2\left(t_D\right)}\right) e^{i\left(p_0 L_D - E_0 t_D\right)/\hbar}, \tag{15}$$

with

$$\tilde{\sigma}_p^2\left(t_D\right) = \sigma_{p_0}^2 \left(1 + i\, t_D \big/ t_{R_\parallel}\right)^{-1},$$

$$\sigma_{p_0} = \hbar \big/ 2\sigma_{z_0}, \tag{16}$$

$$t_{R_\parallel} = \frac{m^*\hbar}{2\sigma_{p_0}^2} = 4\pi\frac{\sigma_{z_0}^2}{\lambda_c^* c},$$

where we define

$$\lambda_c^* = \lambda_c \big/ \gamma^3, \tag{17}$$

with $\lambda_c = h/mc$ – the Compton wavelength. The free-space time development of the wavepacket (15) in the space dimension is calculated by performing the integral over p in Equation (13) with the complex gaussian (15) and the second order expansion of $E_p$ (Eq. 14). Using standard Fourier transform equations for a gaussian:

$$\psi^{(0)}\left(z,t\right) = \frac{\sqrt{\sigma_{z_0}}}{\left(2\pi\tilde{\sigma}_z^4\left(t+t_D\right)\right)^{1/4}} \exp\left(-\frac{\left(z-v_0 t\right)^2}{4\tilde{\sigma}_z^2\left(t+t_D\right)}\right) e^{i\left(p_0\left(z+L_D\right)-\varepsilon_0\left(t+t_D\right)\right)/\hbar}, \tag{18}$$

where t=0 is the entrance time of the center of the wavepacket to the coordinate origin z=0 (later the entrance to the interaction region) after a drift time $t_D$. The complex wavepacket size is defined by $\tilde{\sigma}_z\left(t+t_D\right) = \sigma_{z_0}\sqrt{1+i\left(t+t_D\right)/t_{R_\parallel}}$ . The wavepacket probability distribution

$$\left|\psi^{(0)}\left(z,t\right)\right|^2 = \frac{1}{\sqrt{2\pi\sigma_z^2\left(t+t_D\right)}} \exp\left(-\frac{\left(z-v_0 t\right)^2}{2\sigma_z^2\left(t+t_D\right)}\right), \tag{19}$$

displays "particle-like" propagation at velocity $v_0 = \beta_0 c = p_0/\gamma_0 m$ with wavepacket expansion:

$$\sigma_z(t) = |\tilde{\sigma}_z(t)| = \sigma_{z_0}\sqrt{1 + t^2/t_{R_\parallel}^2} \quad . \tag{20}$$

The parameter $t_{R_\parallel}$ is the evolution time from the waist for which $\sigma_z(t_{R_\parallel}) = \sqrt{2}\sigma_{z0}$, in analogy to the Rayleigh length of wave diffraction.

We now solve Eq. 12 using the first order perturbation theory in momentum space

$$\psi(z,t) = \psi^{(0)}(z,t) + \psi^{(1)}(z,t) = \int dp \left(c_p^{(0)} + c_p^{(1)}\right) e^{-iE_p t/\hbar} |p(z)\rangle. \tag{21}$$

By solving the time-dependent perturbative equation, the finial coefficient obeys the following formula in first-order approximation,

$$i\hbar \dot{c}_{p'}^{(1)} = \int dp\, c_p^{(0)} \langle p'(z)|H_I(t)|p(z)\rangle e^{-i(E_p - E_{p'})t/\hbar}$$

Integration of Eq.12 over time (for $t \to \infty$), produces for $c_p^{(1)}$ two energy conseving delta-function terms of single photon emission/absorption, corresponding to the two terms $\tilde{\mathbf{E}}(z)$ $and$ $\tilde{\mathbf{E}}^*(z)$ in Eq. 10 when used in the perturbation Hamiltonian (eq. 9):

$$c_{p'}^{(1)} = c_{p'}^{(1)(e)} + c_{p'}^{(1)(a)}$$
$$c_{p'}^{(1)(e,a)} = \frac{\pi}{2i\hbar} \int dp \langle p'(z)|H_I^{(e,a)}|p(z)\rangle c_p^{(0)} \delta\left(\frac{E_p - E_{p'} \mp \hbar\omega}{2\hbar}\right), \tag{22}$$

where here and hereafter (e) corresponds to the upper sign, and (a) corresponds to the lower sign. Expanding again the energy dispersion relation (eq.12) to second order

$$E_p = c\sqrt{m^2 c^2 + p^2} \approx \varepsilon_0 + v_0(p - p_0) + \frac{(p - p_0)^2}{2m^*} \quad ,$$ the delta function determines the quantum momentum recoil

$$p_{rec}^{(e,a)} = \left| p'^{(e,a)} - p \right| = \frac{\hbar\omega}{v_0}\left(1 \pm \delta\right),$$

$$\delta = \frac{\hbar\omega}{2m^*v_0^2} \tag{23}$$

The first order perturbation momentum component is then

$$c_{p'}^{(1)(e,a)} = \frac{\pi}{iv_0}\left\langle p'(z)\left|H_I^{(e,a)}\right|\left(p \mp p_{rec}^{(e,a)}\right)(z)\right\rangle c^{(0)}\left(p \mp p_{rec}^{(e,a)}\right), \tag{24}$$

where the zero-order coefficient $c_p^{(0)}$ is given in equation (15). The matrix element is calculated using the perturbation Hamiltonian (9) with the field component (10).

$$\left\langle p'\left|H_I^{(e,a)}\right|p\right\rangle = \int \frac{dz}{2\pi\hbar} H_I^{(e,a)}(0) e^{i(p-p')z/\hbar}$$

$$= \pm \frac{ieE_0 L_I p}{2\pi\gamma_0 m\hbar\omega} sinc\left(\frac{\left(p - p' \mp \hbar q_z\right)L_I}{2\hbar}\right) e^{i(p-p'\mp\hbar q_z)z/2\hbar} e^{\pm i\phi_0}, \tag{25}$$

and then simplified to

$$\left\langle p'\left|H_I^{(e,a)}\right|p \mp p_{rec}^{(e,a)}\right\rangle = \pm\left(\frac{iv_0}{\pi}\right)\Upsilon\left(\frac{p' \mp p_{rec}^{(e,a)}/2}{p_0}\right) sinc\left(\frac{\overline{\theta}_{e,a}}{2}\right) e^{i\frac{\overline{\theta}_{e,a}}{2} \mp i\phi_0}, \tag{26}$$

and

$$c_{p'}^{(1)(e,a)} = \pm\Upsilon\left(\frac{p' \mp p_{rec}^{(e,a)}/2}{p_0}\right) c^{(0)}\left(p' \mp p_{rec}^{(e,a)}\right) sinc\left(\frac{\overline{\theta}_{e,a}}{2}\right) e^{i\frac{\overline{\theta}_{e,a}}{2} \mp i\phi_0}, \tag{27}$$

with

$$\Upsilon = \frac{eE_0 L_I}{2\hbar\omega},$$

$$\overline{\theta}_{e,a} = \overline{\theta} \pm \frac{\varepsilon}{2}, \tag{28}$$

where $\bar{\theta} = \left( \dfrac{\omega}{v_0} - q_z \right) L_I$ is the classical "interaction detuning parameter"[ref.9] and

$\varepsilon = \delta \left( \dfrac{\omega}{v_0} \right) L_I = 2\pi \delta L_I / \beta_0 \lambda$ is the interaction–length quantum recoil parameter[ref. 13 in the article].

### 3. First and second order momentum density distribution and mometum transfer

We now can calculate the electron momentum density distribution after interaction:

$$\rho(p') = \rho^{(0)}(p') + \rho^{(1)}(p') + \rho^{(2)}(p')$$
$$= \frac{\left| c^{(0)}(p') \right|^2 + 2\operatorname{Re}\left\{ c^{(1)*}(p') c^{(0)}(p') \right\} + \left| c^{(1)}(p') \right|^2}{\int dp' \left( \left| c^{(0)}(p') \right|^2 + \left| c^{(1)}(p') \right|^2 \right)}, \qquad (29)$$

where the initial Gaussian momentum density distribution

$$\rho^{(0)}(p') = \left| c^{(0)}(p') \right|^2 = \left( 2\pi\sigma_{p_0}^2 \right)^{-\frac{1}{2}} \exp\left( -\frac{(p-p_0)^2}{2\sigma_{p_0}^2} \right), \qquad (30)$$

is unity normalized - $\int dp' \left| c^{(0)}(p') \right|^2 = 1$.

Using Eq.27, the first order (in terms of $E_0$) momentum-density-distribution is derived from the second term in (29), using (27), (15) :

$$\rho^{(1)}(p') = \frac{2\operatorname{Re}\left\{ c_{p'}^{(1)(\epsilon)*} c_{p'}^{(0)} + c_{p'}^{(1)(a)*} c_{p'}^{(0)} \right\}}{\int dp' \left( \left| c^{(0)}(p') \right|^2 + \left| c^{(1)}(p') \right|^2 \right)} = \frac{2\Upsilon}{1 + \Upsilon^2 \left( sinc\left( \dfrac{\bar{\theta}_e}{2} \right) + sinc\left( \dfrac{\bar{\theta}_a}{2} \right) \right)} \times$$

$$\operatorname{Re}\left\{ \left( \frac{p' - p_{rec}^e/2}{p_0} \right) \left( 2\pi\sigma_{p0}^2 \right)^{-\frac{1}{2}} e^{-\frac{(p'-p_0)^2}{4\hat{\sigma}_p^2(t_D)} - \frac{(p'-p_0-p_{rec}^e)^2}{4\hat{\sigma}_p^2(t_D)}} sinc\left( \frac{\bar{\theta}_e}{2} \right) e^{i\left( \frac{\bar{\theta}_e}{2} + \phi_0 \right)} \right.$$

$$\left. - \left( \frac{p' + p_{rec}^a/2}{p_0} \right) \left( 2\pi\sigma_{p0}^2 \right)^{-\frac{1}{2}} e^{-\frac{(p'-p_0)^2}{4\hat{\sigma}_p^2(t_D)} - \frac{(p'-p_0+p_{rec}^a)^2}{4\hat{\sigma}_p^2(t_D)}} sinc\left( \frac{\bar{\theta}_a}{2} \right) e^{i\left( \frac{\bar{\theta}_a}{2} - \phi_0 \right)} \right\}, \qquad (31)$$

This apparently quantum regime expression depends on the quantum recoil, the synchronizm condition and also the acceleration phase. For the purpose of comparison with the classical point-particle acceleration (that is maximal near synchronism - $\bar{\theta} \simeq 0$), we consider the case of $\varepsilon \ll 1$ ( short interaction length) and substitute $\bar{\theta}_e \simeq \bar{\theta}_a \simeq \bar{\theta}$. This simplifies the expression to:

$$\rho^{(1)}(p') = \frac{2\pi \Upsilon e^{-\Gamma^2/2} sinc\left(\dfrac{\bar{\theta}}{2}\right)}{1 + 2\Upsilon^2 sinc^2\left(\dfrac{\bar{\theta}}{2}\right)} cos\left(\frac{\bar{\theta}}{2} + \phi_0 + \varphi\right)\left|C_{p'}^{(0)}\right|^2$$

$$\left((p' - p_{rec}^e/2)e^{-\left(\frac{p'-p_0}{2\sigma_{p_0}^2}\right)p_{rec}^e + \delta\Gamma^2} - (p' + p_{rec}^a/2)e^{\left(\frac{p'-p_0}{2\sigma_{p_0}^2}\right)p_{rec}^a - \delta\Gamma^2}\right),$$

(32)

where

$$\varphi = \frac{2\omega t_D}{m v_0}(p' - p_0),$$

$$\Gamma = \frac{\omega}{v_0}\sigma_z(t_D) = \frac{2\pi\sigma_z(t_D)}{\beta\lambda}.$$

(33)

The decay coefficient $e^{-\Gamma^2/2}$ assures that the first order term of the momentum distribution $\rho^{(1)}$ is diminished when $\Gamma \gg 1$, namely when the wavepacket expands on its way from the source beyond the size of the interaction wavelength. Eq.31 can be furthur simplified by first order expansion in $\delta \ll 1$, resulting in

$$\rho^{(1)}(p') = 2\left(\frac{\hbar\omega/v_0}{p_0}\right)\Upsilon\rho^{(0)}(p') sinc\left(\frac{\bar{\theta}}{2}\right)e^{-\Gamma^2/2}cos\left(\frac{\bar{\theta}}{2} + \phi_0 + \varphi\right)\left[1 - \frac{(p'-p_0)p'}{\sigma_{p_0}^2}\right]. \quad (34)$$

and assuming $\varphi \ll 1$

$$\Delta p^{(1)} = \int \rho^{(1)}(p')p'dp' = \left(-\frac{eE_0 L_l}{v_0}sinc\left(\frac{\bar{\theta}}{2}\right)cos\left(\phi_0 + \frac{\bar{\theta}}{2}\right)\right)e^{-\Gamma^2/2} = \Delta p_{point}e^{-\Gamma^2/2}, \quad (35)$$

Where $\Delta p_{point}$ is the classical point-particle momentum transfer (11).

The second order (in terms of $E_0$) momentum-density-distribution is derived from the third term in Eq.29 (second order),and the same order perturbation contribution from the expansion of the denomintor:

$$\rho^{(2)}(p') \simeq \left|c^{(1)}(p')\right|^2 - \rho^{(0)}(p')\int dp'\left|c^{(1)}(p')\right|^2 \tag{36}$$

Neglecting the interference between the emission and absorption terms $\left|c_{p'}^{(1)(e)} + c_{p'}^{(1)(a)}\right|^2$, and using (27) one obtains:

$$\rho^{(2)}(p') \simeq \left(\left|c^{(1)(e)}(p')\right|^2 + \left|c^{(1)(a)}(p')\right|^2\right) - \rho^{(0)}(p')\int dp'\left(\left|c^{(1)(e)}(p')\right|^2 + \left|c^{(1)(a)}(p')\right|^2\right) =$$

$$\Upsilon^2\left[\left(\frac{p' + p_{rec}^e/2}{p_0}\right)^2 \rho^{(0)}\left(p' + p_{rec}^e\right) - \left(\frac{p_0 - p_{rec}^{(e)}/2}{p_0}\right)^2 \rho^{(0)}(p')\right]sinc^2\left(\frac{\bar{\theta}_e}{2}\right) \tag{37}$$

$$+ \Upsilon^2\left[\left(\frac{p' - p_{rec}^a/2}{p_0}\right)^2 \rho^{(0)}\left(p' - p_{rec}^a\right) - \left(\frac{p_0 + p_{rec}^{(a)}/2}{p_0}\right)^2 \rho^{(0)}(p')\right]sinc^2\left(\frac{\bar{\theta}_a}{2}\right).$$

The second order momentum transfer is therefore

$$\Delta p^{(2)} = \int \rho^{(2)}(p')\, p' dp'$$

$$= \Upsilon^2 sinc^2\left(\frac{\bar{\theta}_e}{2}\right)\int\left[\left(\frac{p' + p_{rec}^e/2}{p_0}\right)^2 \rho^{(0)}\left(p' + p_{rec}^e\right) - \left(\frac{p_0 - p_{rec}^{(e)}/2}{p_0}\right)^2 \rho^{(0)}(p')\right].p' dp' +$$

$$\Upsilon^2 sinc^2\left(\frac{\bar{\theta}_a}{2}\right)\int\left[\left(\frac{p' - p_{rec}^a/2}{p_0}\right)^2 \rho^{(0)}\left(p' - p_{rec}^a\right) - \left(\frac{p_0 + p_{rec}^{(a)}/2}{p_0}\right)^2 \rho^{(0)}(p')\right].p' dp'$$

$$= \Upsilon^2 sinc^2\left(\frac{\bar{\theta}_e}{2}\right)\left(\left(p_0 - 2p_{rec}^e\right) - p_{rec}^e\left(\frac{3}{4}\frac{p_{rec}^e}{p_0} + \frac{1}{4}\left(\frac{p_{rec}^e}{p_0}\right)^2\right) + \left(\frac{\sigma_{p_0}}{p_0}\right)^2\left(3p_0 - 2p_{rec}^e\right)\right) + \text{other three terms.} \tag{38}$$

The explicit expression for the first term was obtained by direct (but tedious) integration over p' with $\rho^{(0)}(p')$ taken to be the Gaussian momentum distribution (30). Similar expressions are directly derived from similar integration of the other three terms, and with the approximation, $p_{rec}^{e,a}, \sigma_{p_0} \ll p_0$, this results in:

$$\Delta p^{(2)} \simeq \Upsilon^2 sinc^2\left(\frac{\overline{\theta}_e}{2}\right)\left(\left(p_0 - 2p_{rec}^e\right) - \left(p_0 - p_{rec}^e\right)\right) + \Upsilon^2 sinc^2\left(\frac{\overline{\theta}_a}{2}\right)\left(\left(p_0 + 2p_{rec}^a\right) - \left(p_0 + p_{rec}^a\right)\right)$$

$$= -\Upsilon^2 \left(\frac{\hbar\omega}{v_0}\right)\left[sinc^2\left(\frac{\overline{\theta}_e}{2}\right) - sinc^2\left(\frac{\overline{\theta}_a}{2}\right)\right].$$

$$(39)$$

At this point, it is proper to explain the neglect in Eqs. 36&37 of the phase-dependent second order emission and absorption terms $2\text{Re}\left\{c^{(2)*}\left(p'\right)c^{(0)}\left(p'\right) + c^{(1)(e)*}\left(p'\right)c^{(1)(a)}\left(p'\right)\right\} \propto e^{-4\Gamma^2/2}$ and other second order processes of emission-absorption and absorption-emission. Their inclusion requires second order perturbation analysis [14, 38] beyond the present first order perturbation analysis. Their inclusion would not affect the main results of the derived first-order momentum density expressions. They would add small wavepacket-dependent contributions to the second-order momentum density expression (36), and will produce second order sidebands of two-photon emission and two-photon absorption processes in the PINEM spectrum of Fig. 2.

## 4. Measurement limits

It is necessary to realize that convential electron energy spectroscopy necessitates averaged measurement of a multitude of electrons, in order to view the momentum distribution. The momentum distribution of an ensemble of electrons is found from integration over coordinate z of the Wigner distribution for fixed p, in the p-z phase-space [33] (See Fig. 2). For an ensemble of uncorrelated single electron measurements, this reduces to a simple classical statistical averageing integral over the initial (thermal) central momenta distribution with the single electron wavepacket momentum distribution:

$$\rho_{en} = \int \rho\left(p', p_0\right) f\left(\frac{p_0 - p_{0,en}}{\sigma_{p,th}}\right) dp_0,$$

$$(40)$$

This, necessarily, results in widening of the measurable standard deviation of the Gaussian distribution of the ensemble to $\sigma_{p,en}$:

$$\sigma_{p,en}^2 = \sigma_{p,th}^2 + \sigma_{p_0}^2. \qquad (41)$$

Since the measurable distribution satisfies $\sigma_{p,en} > \sigma_{p0}$, the wavepacket broadening cannot be distinguished from the thermal distribution that determines the so called "Coherence time" of electron microscopes $t_{coh} = \hbar / 2\sigma_{E,en}$ (typically $\sigma_{E,en} < \sim 0.7 eV$). However, the quantum recoil effect on the momentum distribution of both emission and absorption is still observable in the post-interaction energy spectrum of the electron $\rho^{(2)}(p')$ if $p_{rec}^{(0)} > \sigma_{p,en}$ as shown in Figure 2, in agreement with PINEM experiments[ref:18-22 in the text].

## 5. Numerical simulation

**Video-1:** Simulation of stimulated interaction in the quantum large recoil limit.

Website: https://www.youtube.com/watch?v=dEpsjHgL2QY

Wavepacket evolution for the case $\bar{\theta} = 0$ and weak field $\Upsilon \ll 1, \sigma_{p0} < \hbar\omega / v_0, \sigma_z(t_D) > \lambda\beta_0$, with sidebands formation and no net acceleration: simulation parameters $\beta_0 = 0.7$, $\Gamma = 3$, $\Upsilon = 0.2$, $\phi_0 = 0$

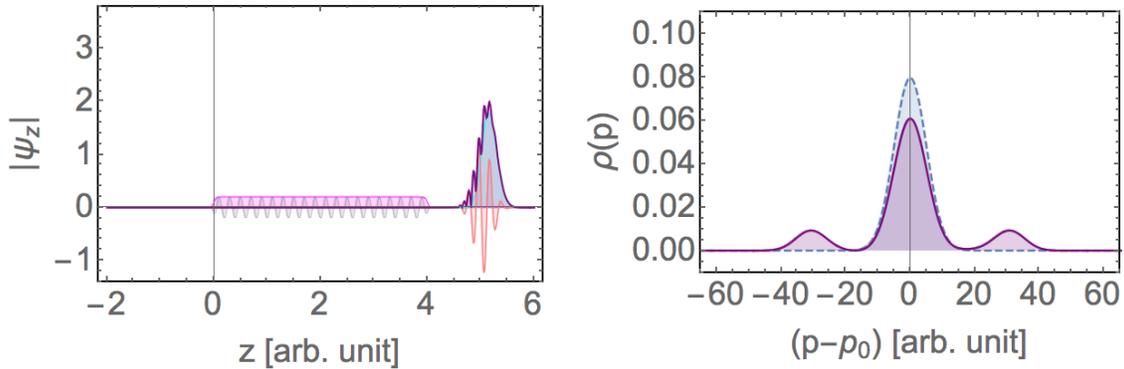

(Left) Evolution in space -z (pink- $\operatorname{Im}\Psi(z)$ , cyan- $\operatorname{Re}\Psi(z)$ and purple- $|\Psi(z)|$ ), the central pink area is the near-field radiation amplitude and phase. (Right) Evolution of momentum density $\rho(p)=|c(p)|^2$ is symetric without net momentum gain.

**Video-2:** Simulation of phase-dependent stimulated-interaction in the quantum-classical wavepacket transition regime.

Website: https://www.youtube.com/watch?v=BE5oJBmwNyc

Wavepacket evolution for the case $\bar{\theta}=0$ and <u>weak field</u> $\Upsilon \ll 1, \sigma_z(t_D)<\lambda\beta_0$ , with simulation parameters: $\beta_0=0.7$, $\Gamma=0.6$, $\Upsilon=0.2$. (a) $\phi_0=0$ (acceleration); (b) $\phi_0=\pi$ (decceleration):

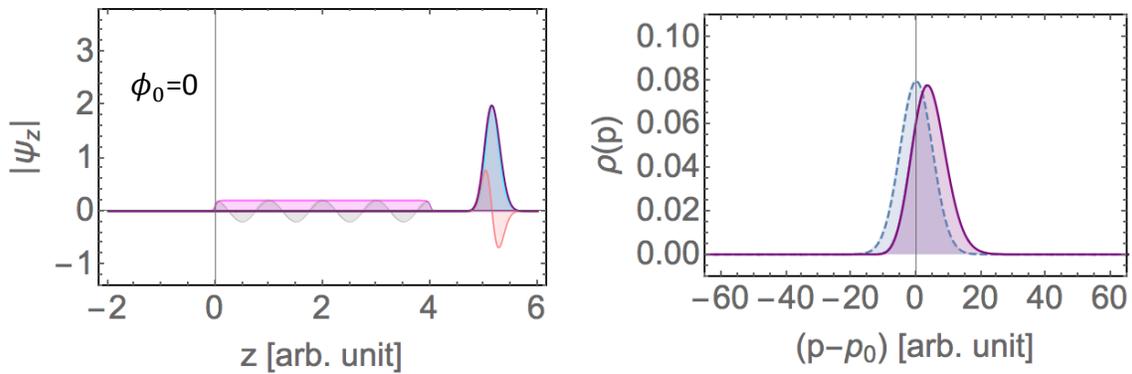

a. (Left) Evolution in space- z. (Right) Evolution of momentum density distribution $\rho(p)=|c(p)|^2$ with net <u>positive</u> momentum gain.

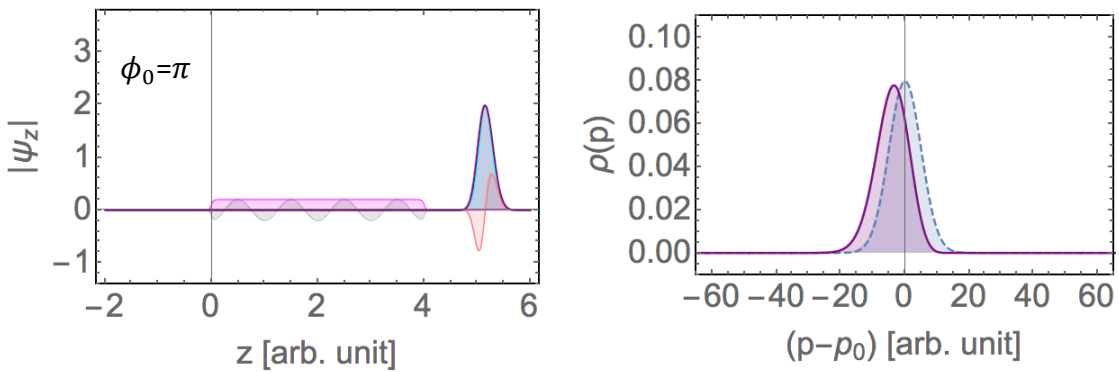

b. (Left) Evolution in space- z. (Right) Evolution of momentum density distribution $\rho(p) = |c(p)|^2$ with net <u>negative</u> momentum gain.

**Video-3:** Simulation of phase-dependent interaction in the near "point-particle" limit.

Website: https://www.youtube.com/watch?v=f5gWvAd4QRI

Wavepacket evolution for the case $\bar{\theta} = 0$ and <u>strong field</u> (multiphoton exchange) $\Upsilon > 1$, $\sigma_z(t_D) < \lambda\beta_0$ , with net acceleration/deceleration simulation parameters: $\beta_0 = 0.7$, $\Gamma = 0.6$, $\Upsilon = 1.5$. (a) $\phi_0 = 0$ (acceleration);(b) $\phi_0 = \pi$ (deceleration):

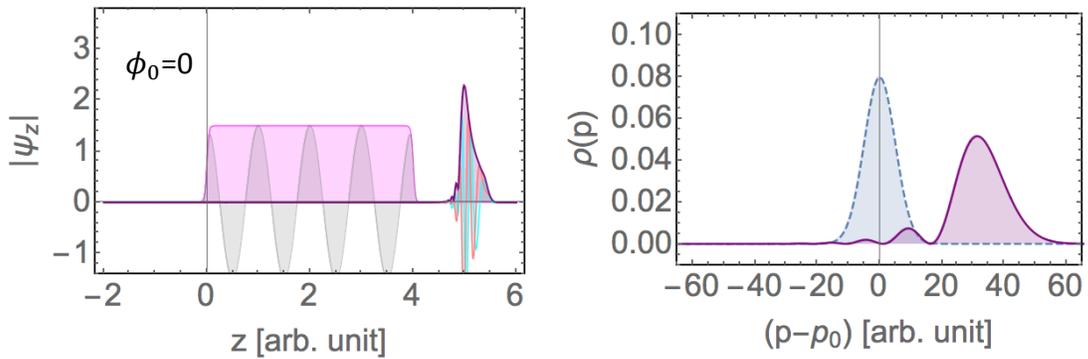

a. (Left) Evolution in space- z. (Right) Evolution of momentum density distribution $\rho(p) = |c(p)|^2$ with net <u>positive</u> momentum gain.

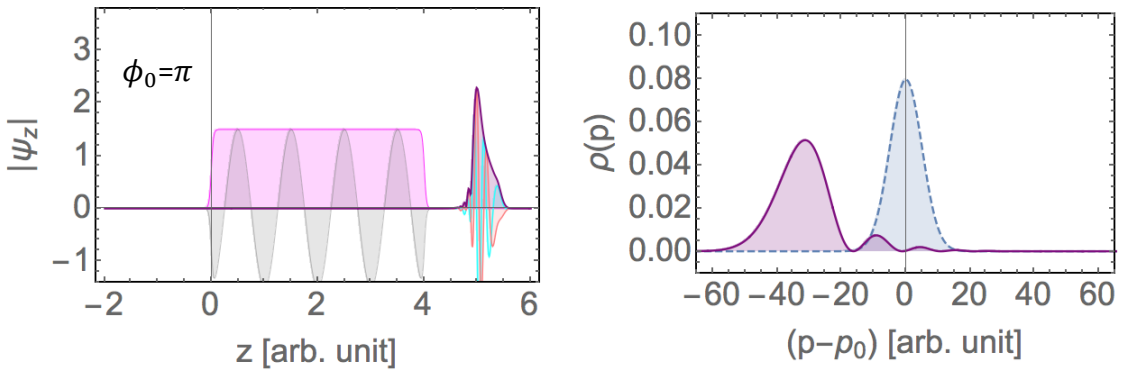

b. (Left) Evolution in space- z. (Right) Evolution of momentum density distribution $\rho(p) = |c(p)|^2$ with net <u>negative</u> momentum gain.